\documentclass[letter,traditabstract]{aa}
\usepackage{amssymb}
\usepackage{graphics}
\usepackage{tabularx}
\usepackage{color}
\usepackage{natbib}
\usepackage{rotating}

\def\ek{2010\,EK\ensuremath{_{139}}}
\def\sedna{(90377)~Sedna}

\def\refmark#1{#1}

\begin{document}


\title{``TNOs are Cool'': A survey of the trans-Neptunian region }

\subtitle{VII. Size and surface characteristics of \sedna{} and \ek{}}

\author{%
Andr\'as P\'al\inst{1,2} \and
Csaba Kiss\inst{1} \and
Thomas G. M\"uller\inst{3} \and
Pablo Santos-Sanz\inst{4} \and
Esa Vilenius\inst{3} \and
Nikolett Szalai\inst{1} \and
Michael Mommert\inst{5} \and
Emmanuel Lellouch\inst{4} \and
Miriam Rengel\inst{6} \and
Paul Hartogh\inst{6} \and
Silvia Protopapa\inst{7,6} \and
John Stansberry\inst{8} \and
Jose-Luis Ortiz\inst{9} \and
Ren\'e Duffard\inst{9} \and
Audrey Thirouin\inst{9} \and
Florence Henry\inst{4} \and
Audrey Delsanti\inst{4,10}}

\institute{%
Konkoly Observatory, MTA Research Centre for Astronomy and Earth Sciences,
	Konkoly-Thege Mikl\'os \'ut 15-17,
	1121 Budapest, Hungary; e-mail: \texttt{apal@szofi.net} \and
Department of Astronomy, Lor\'and E\"otv\"os University, 
	P\'azm\'any P\'eter s\'et\'any 1/A, 
	1117 Budapest, Hungary \and
Max-Planck-Institut f\"ur extraterrestrische Physik, 
	Postfach 1312, Giessenbachstr., 85741 Garching, Germany \and
LESIA-Observatoire de Paris, CNRS, UPMC Univ. Paris 06, Univ. Paris-Diderot,
	5 Place J. Janssen, 92195 Meudon Cedex, France \and
Deutsches Zentrum f\"ur Luft- und Raumfahrt (DLR), Institut f\"ur Planetenforschung, 
	Rutherfordstrasse 2, 12489 Berlin, Germany \and
Max-Planck-Institut f\"ur Sonnensystemforschung, 
	Max-Planck-Strasse 2, 
	37191 Katlenburg-Lindau, Germany \and
Department of Astronomy, University of Maryland,
	College Park, MD 20742-2421, USA \and
Steward Observatory, University of Arizona, 
	933 North Cherry Avenue, Tucson, AZ 85721, USA \and
Instituto de Astrof\'{\i}sica de Andaluc\'{\i}a - CSIC, 
	Apt 3004, 18008 Granada, Spain \and
Laboratoire d'Astrophysique de Marseille, CNRS \& Universit\'e de Provence, 
	38 rue Fr\'ed\'eric Joliot-Curie, 13388 Marseille cedex 13, France
}


\date{Received \dots; accepted \dots}

\abstract{%
We present estimates of the basic physical properties 
(size and albedo) of \sedna{}, a prominent member of the detached 
trans-Neptunian object population and 
the recently discovered scattered disk object \ek{}, based 
on the recent observations acquired with the 
Herschel Space Observatory\thanks{{\it Herschel} is an ESA space 
observatory with science instruments provided by European-led 
Principal Investigator consortia and with important participation from NASA.},
within the ``TNOs are Cool!'' key programme. 
Our modeling of the thermal measurements shows that both objects
have larger albedos and smaller sizes than the previous expectations,
thus their surfaces might be covered by ices in a significantly
larger fraction. The derived diameter of Sedna and 
\ek{} are $995\pm80$\,km and $470^{+35}_{-10}\,{\rm km}$, while the
respective geometric albedos are $p_{\rm V}\!=\!0.32\pm0.06$ and 
$0.25^{+0.02}_{-0.05}$. These estimates are
based on thermophysical model techniques.}


\maketitle


\section{Introduction}

The Herschel Space Observatory \citep{pilbratt2010} allows
the detection of thermal radiation from 
\refmark{several} trans-Neptunian objects (TNOs) 
at the precision level of $<1\,{\rm mJy}$. 
Since the expected fluxes around the peak of the spectral
energy distribution (SED) significantly exceed this precision,
\emph{Herschel} provides a great opportunity to characterize TNOs
and obtain basic thermophysical information. 
In this work, we present recent measurements 
of the prominent objects \sedna{} and \ek{} 
using
the Photodetector Array Camera and Spectrometer
instrument \citep[PACS][]{poglitsch2010} on board 
the Herschel Space Observatory. These observations are part 
of the ``TNOs are Cool!: a survey of the trans-Neptunian region'' Open Time
Key Program \citep{mueller2009,mueller2010,lellouch2010,lim2010}.

Sedna is a prominent member of the detached 
objects, that is often classified as an inner Oort-cloud object.
Until now, no accurate measurements of the diameter and albedo
have been available for this object. Both direct imaging \citep{brown2008} and
upper limits to the thermal radiation using the 
Spitzer Space Telescope \citep[$2.4\,{\rm mJy}$ at $70\,\mu{\rm m}$, see][]{stansberry2008}
have yielded an upper limit of $\approx 1670\,{\rm km}$ for its diameter
\refmark{(within 97\% confidence)}.

\ek{} has been discovered in 2010 by \cite{sheppard2011} in the 
course of a southern Galactic plane survey. 
Prediscovery observations date back to 2002, allowing for a relatively
accurate orbit determination. This places \ek{} among the scattered disk
objects. \ek{} orbits the Sun on an eccentric orbit ($e\approx0.53$)
and has a perihelion distance of $q\approx 32.5\,{\rm AU}$. In addition,
\ek{} is \refmark{a suspected member of the 2:7 resonance}
group\footnote{http://boulder.swri.edu/\~{ }buie/kbo/astrom/10EK139.html}.
We note that a more complete sample of SDOs/detached objects
observed with Herschel/PACS is presented by \cite{santossanz2012}.


\begin{table}
\caption{Summary of Herschel observations of Sedna 
and \ek{}. The columns are: 
i) observation identifier, 
ii) date and time, 
iii) scan angle direction with respect to the detector array, and 
iv) filter configuration.}
\label{table:herschelobs}
\begin{center}\begin{tabular}{lrrl}
\hline
OBSID & Date \& Time (UT) & Angle & Filter \\
\hline
Sedna \\
\hline
1342202227 & 2010-08-06 10:55:17 & $ 70^\circ$ & B/R \\
1342202228 & 2010-08-06 11:19:54 & $110^\circ$ & B/R \\
1342202229 & 2010-08-06 11:44:31 & $ 70^\circ$ & G/R \\
1342202230 & 2010-08-06 12:09:08 & $110^\circ$ & G/R \\
1342202306 & 2010-08-09 08:11:37 & $ 70^\circ$ & B/R \\
1342202307 & 2010-08-09 08:36:14 & $110^\circ$ & B/R \\
1342202308 & 2010-08-09 09:00:51 & $ 70^\circ$ & G/R \\
1342202309 & 2010-08-09 09:25:28 & $110^\circ$ & G/R \\
\hline
\ek{} \\
\hline
1342211418 & 2010-12-23 07:04:30 & $ 70^\circ$ & B/R \\
1342211419 & 2010-12-23 07:15:01 & $110^\circ$ & B/R \\
1342211420 & 2010-12-23 07:25:32 & $ 70^\circ$ & G/R \\
1342211421 & 2010-12-23 07:36:03 & $110^\circ$ & G/R \\
1342211524 & 2010-12-23 19:58:27 & $ 70^\circ$ & B/R \\
1342211525 & 2010-12-23 20:08:58 & $110^\circ$ & B/R \\
1342211526 & 2010-12-23 20:19:29 & $ 70^\circ$ & G/R \\
1342211527 & 2010-12-23 20:30:00 & $110^\circ$ & G/R \\
\hline
\end{tabular}\end{center}\vspace*{-3mm}
\end{table}

\section{Observations, data reduction and photometry}
\label{sec:observations}


Sedna was observed by Herschel/PACS
in two visits: the first started on 
2010 August 6, 10:55:17 UTC
and a follow-up started on
2010 August 9, 08:11:37 UTC, 
both taking place during the Routine Science Phase 
observation series of the ``TNO's are Cool!'' key 
programme \citep{mueller2009}. \ek{} was also observed by Herschel/PACS
in two visits, the first started on
2010 December 23, 07:04:30 UTC,
and a follow-up started the same day, 19:58:27 UTC.
Herschel/PACS observed Sedna and \ek{} for 
$\approx 3.14$ and $\approx 1.26$ hours, respectively. For both objects, we used
both the blue/red ($70/160\,\mu{\rm m}$) and green/red 
($100/160\,\mu{\rm m}$) channel combinations.
The actual details of these observations are summarized in 
Table~\ref{table:herschelobs}.


Raw observational data were reduced using the
Herschel Interactive Processing
Environment (HIPE\footnote{Data presented in this paper were analyzed using
``HIPE'', a joint development by the Herschel Science Ground Segment
Consortium, consisting of ESA, the NASA Herschel Science Center, and the HIFI,
PACS and SPIRE consortia members, see
http://herschel.esac.esa.int/DpHipeContributors.shtml.}, see also Ott, 2010)
and the processing scripts are similar to the ones employed in \cite{mommert2012},
\cite{santossanz2012}, or \cite{vilenius2012}.
For each observation, these scripts create a pair of maps, one
for the blue or green channel and one for the red channel. 
The maps have an effective pixel size of $1.\!\!^{\prime\prime}1$, 
$1.\!\!^{\prime\prime}4$, and $2.\!\!^{\prime\prime}1$, 
for the blue, green, and red filters, respectively: these pixel sizes 
are set to sample the respective point spread functions (PSFs) properly.
Data frames were selected by the actual scan speed
($10^{\prime\prime}/{\rm sec}\le{\rm speed}\le 25^{\prime\prime}/{\rm sec}$)
of the spacecraft, which maximized the effective usage of the 
detector and yielded significantly higher signal-to-noise (S/N) ratios 
than the standard setting (approximately $20^{\prime\prime}/{\rm sec}$).

\begin{figure}
\begin{center}
\begin{tabular}{cccc}
& $70\,\mu{\rm m}$ & $100\,\mu{\rm m}$ & $160\,\mu{\rm m}$ \\
\begin{sideways}\hspace*{10mm}Sedna\end{sideways} & 
\hspace*{-4mm}\resizebox{30mm}{!}{\includegraphics{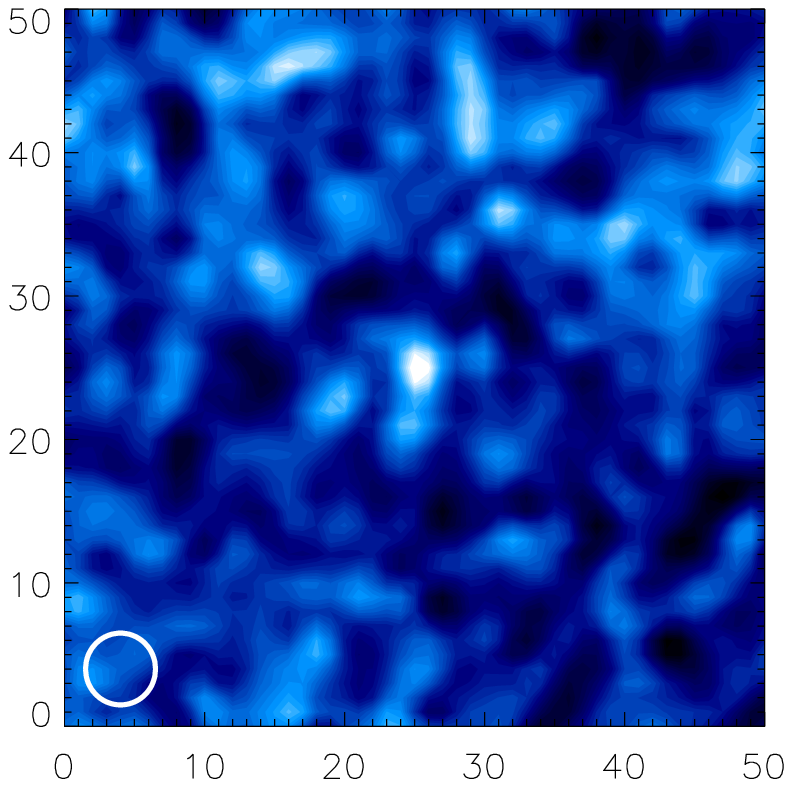}} \hspace*{-4mm}&
\hspace*{-4mm}\resizebox{30mm}{!}{\includegraphics{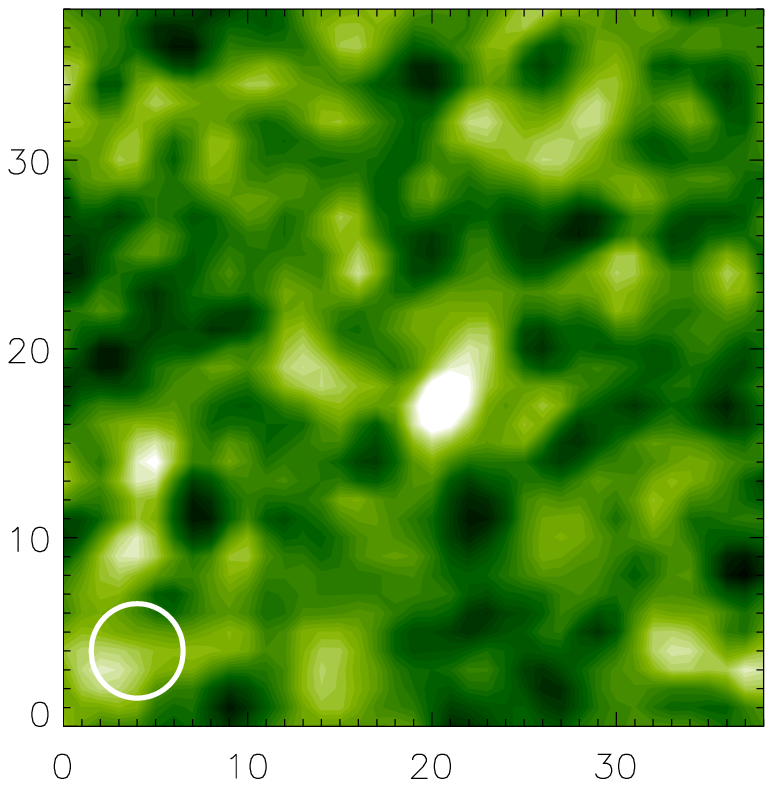}}\hspace*{-4mm} &
\hspace*{-4mm}\resizebox{30mm}{!}{\includegraphics{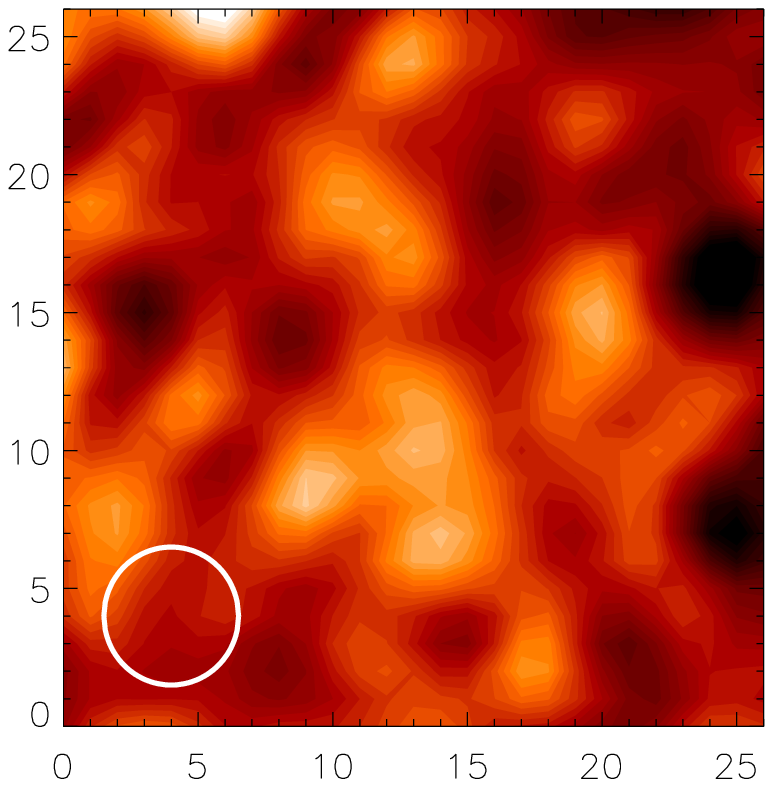}} \\
\begin{sideways}\hspace*{8mm}\ek{}\end{sideways} & 
\hspace*{-4mm}\resizebox{30mm}{!}{\includegraphics{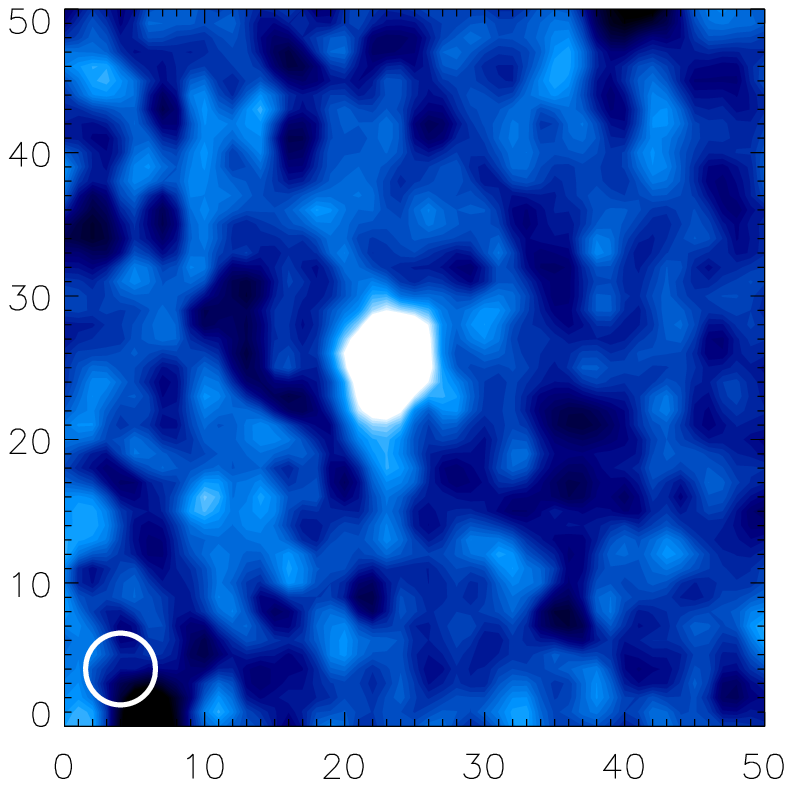}} \hspace*{-4mm}&
\hspace*{-4mm}\resizebox{30mm}{!}{\includegraphics{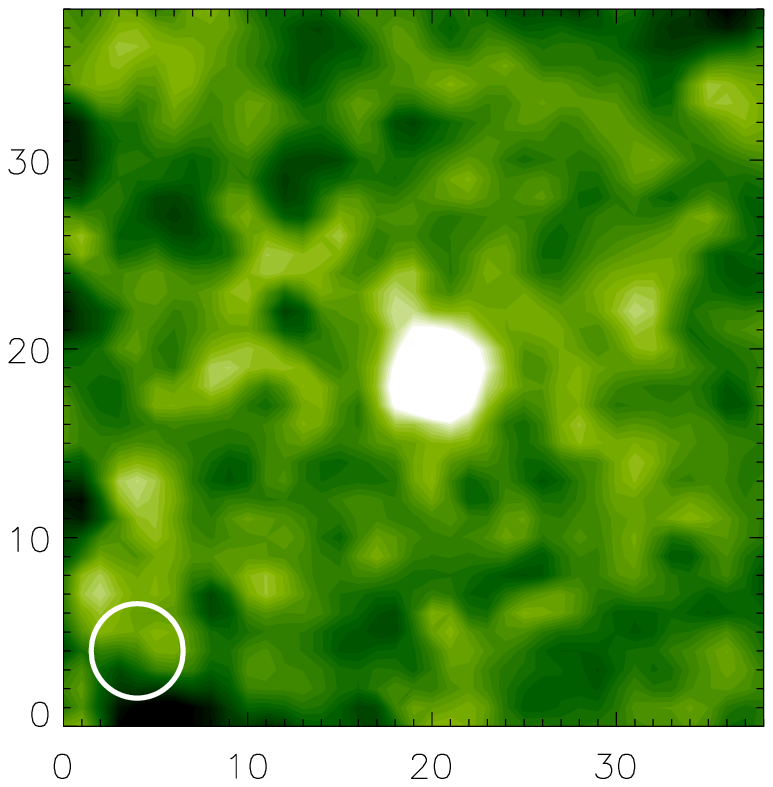}}\hspace*{-4mm} &
\hspace*{-4mm}\resizebox{30mm}{!}{\includegraphics{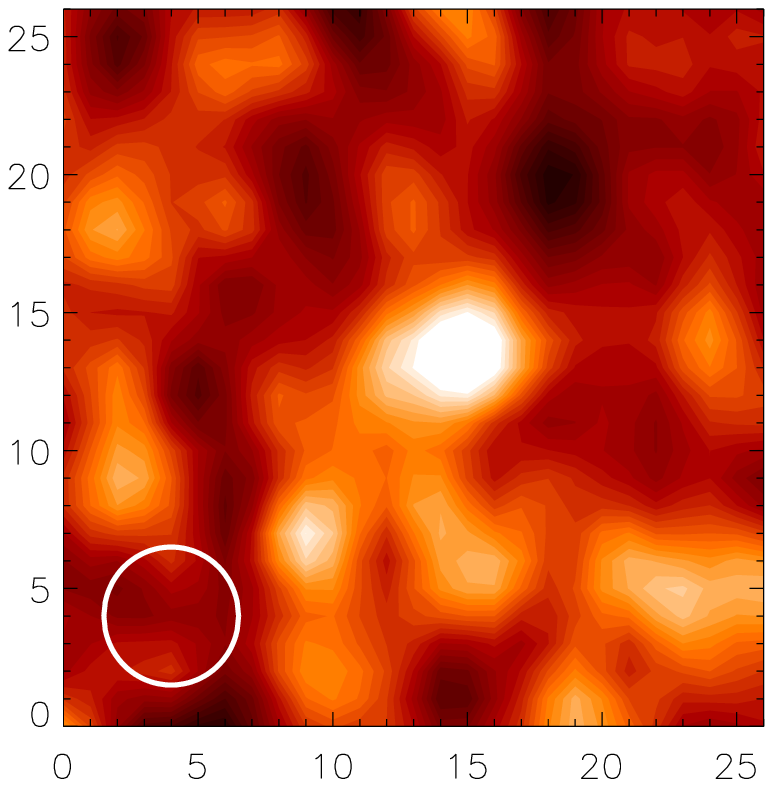}}
\end{tabular}
\end{center}\vspace*{-4mm}
\caption{Image stamps showing the combined maps of Sedna (upper panels) and 
\ek{} (lower panels) in the $70\,\mu{\rm m}$ (blue),
 $100\,\mu{\rm m}$ (green), and $160\,\mu{\rm m}$ (red) channels.
Each stamp covers an area of $56^{\prime\prime}\times56^{\prime\prime}$, while 
the tick marks on the axes show the 
relative positions in pixels. The effective beam size (i.e. the circle with a
diameter corresponding to the full width at half magnitude) 
is also displayed in the lower-left corners
of the stamps.}
\label{fig:sednaekimagestamps}
\end{figure}

Since the apparent motion of Sedna and \ek{} between the two
visits ($15-35$ map pixels, depending on the actual filter) is
relatively large compared to the PSF but small compared to the detector size,
the location of the target in the first visit can simply
be used as a background area on the maps of the second visit and vice versa.
Owing to the satellite pointing uncertainty that is 
about a few arcsec \citep{poglitsch2010},
we derived the true map-center displacements using the 
red channel maps -- on which the background confusion is
the strongest -- as follows. By varying the 
proper motion vector between the two visits, we computed the 
cross-correlation residuals for each trial vector. By minimizing
the residuals, we obtained a more precise value for
the shift between the visits and the photometry of combined maps 
was found to be more reliable. 
Since simple averaging the registered maps does not cancel the
background confusion noise, \refmark{we employed background removal
techniques as it is described in \cite{santossanz2012} or \cite{mommert2012}.
The maps on which the photometry was then performed 
are shown in Fig.~\ref{fig:sednaekimagestamps}.}

\begin{table}
\caption{Thermal fluxes of Sedna and 
\ek{} derived from our Herschel measurements.} 
\label{table:herscheldata}
\begin{center}\begin{tabular}{llrr}
\hline
Object & Band & $\lambda$ & Flux \\
\hline
Sedna	&  B	& $70\,{\rm\mu m}$	& $1.8\pm0.7\,{\rm mJy}$	\\
	&  G	& $100\,{\rm\mu m}$	& $4.2\pm0.9\,{\rm mJy}$	\\
	&  R	& $160\,{\rm\mu m}$	& $2.7\pm1.3\,{\rm mJy}$	\\
\hline
\ek{}	&  B	& $70\,{\rm\mu m}$	& $17.4\pm1.1\,{\rm mJy}$	\\
	&  G	& $100\,{\rm\mu m}$	& $16.3\pm1.4\,{\rm mJy}$	\\
	&  R	& $160\,{\rm\mu m}$	& $11.9\pm1.8\,{\rm mJy}$	\\
\hline
\end{tabular}\end{center}\vspace*{-3mm}
\end{table}

\begin{table}
\caption{Orbital and optical data for Sedna 
and \ek{} at the time of the Herschel observations. The parameters $r$ and 
$\Delta$ denote the heliocentric distance and the 
distance from Herschel, $\alpha$ is the phase angle,
and $H_{\rm V}$ is the absolute visual magnitude, which is available
from the literature.} 
\label{table:auxdata}
\begin{center}\begin{tabular}{llr}
\hline
Object	& Quantity	& Value \\
\hline
Sedna	& $r$		& $87.43$\,AU		\\
	& $\Delta$ 	& $87.56$\,AU		\\
	& $\alpha$	& $0.\!\!^\circ7$	\\
	& $H_{\rm V}$ 	& $+1.83\pm 0.05$	\\
\hline
\ek{}	& $r$		& $39.08$\,AU		\\
	& $\Delta$ 	& $39.50$\,AU		\\
	& $\alpha$	& $1.\!\!^\circ3$	\\
	& $H_{\rm V}$ 	& $+3.80\pm 0.10$	\\
\hline
\end{tabular}\end{center}\vspace*{-3mm}
\end{table}


\refmark{Regardless of the background structures, in the subtracted
and combined maps the only
expected source is the TNO itself and this source can be treated as an
isolated point source.}
We estimated the fluxes and their uncertainties using (1) a single 
aperture that maximizes the expected S/N ratio;
(2) the aperture growth curve method and implanted artificial sources in 
a Monte Carlo fashion \citep[see e.g.][]{santossanz2012}; 
and (3) we also checked the individual (non-combined)
maps on which they had sufficient S/N ratio. For 
a more detailed description, we refer to \cite{mommert2012} and \cite{santossanz2012}. 

\begin{figure*}
\begin{center}
\resizebox{80mm}{!}{\includegraphics{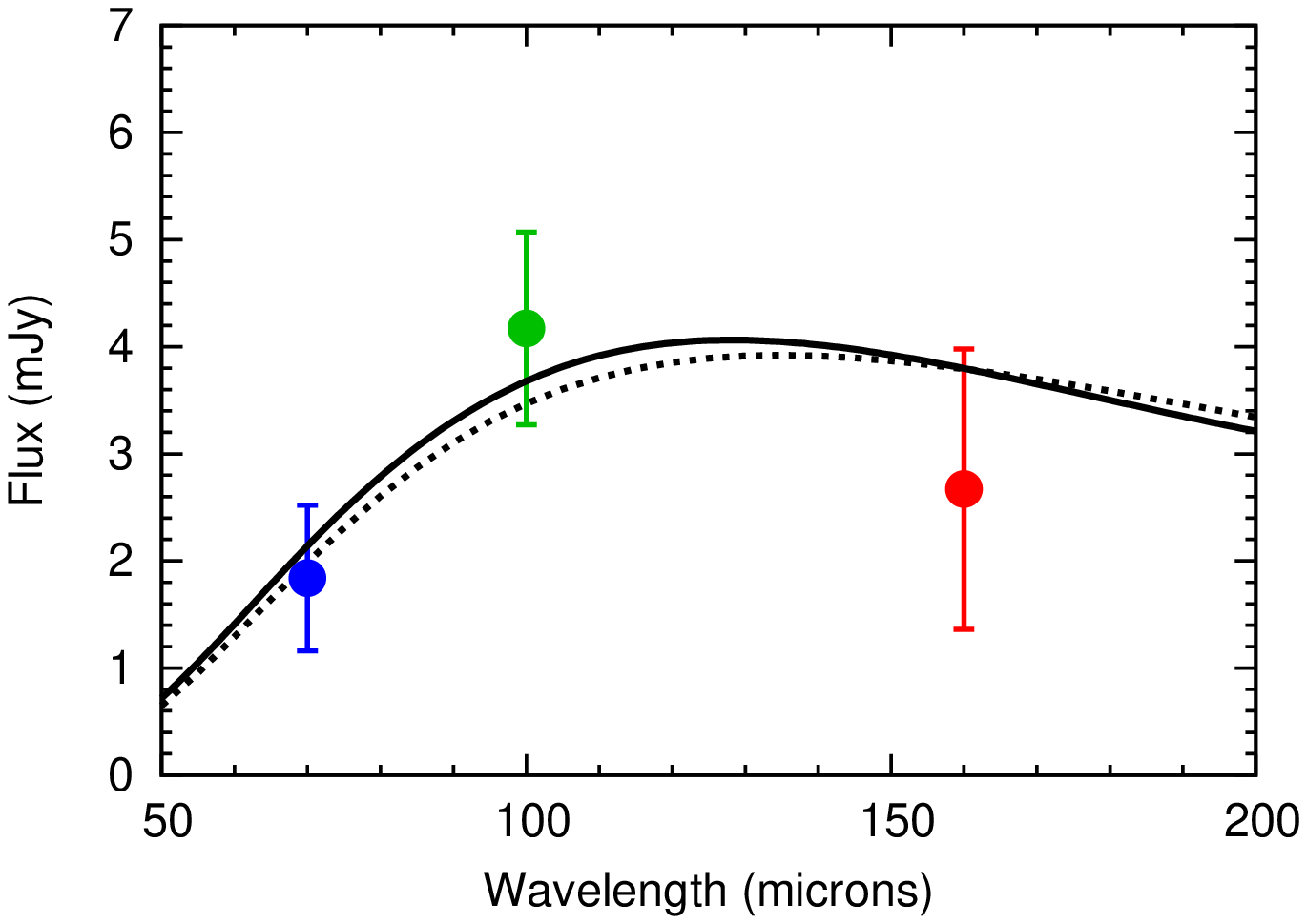}}%
\resizebox{80mm}{!}{\includegraphics{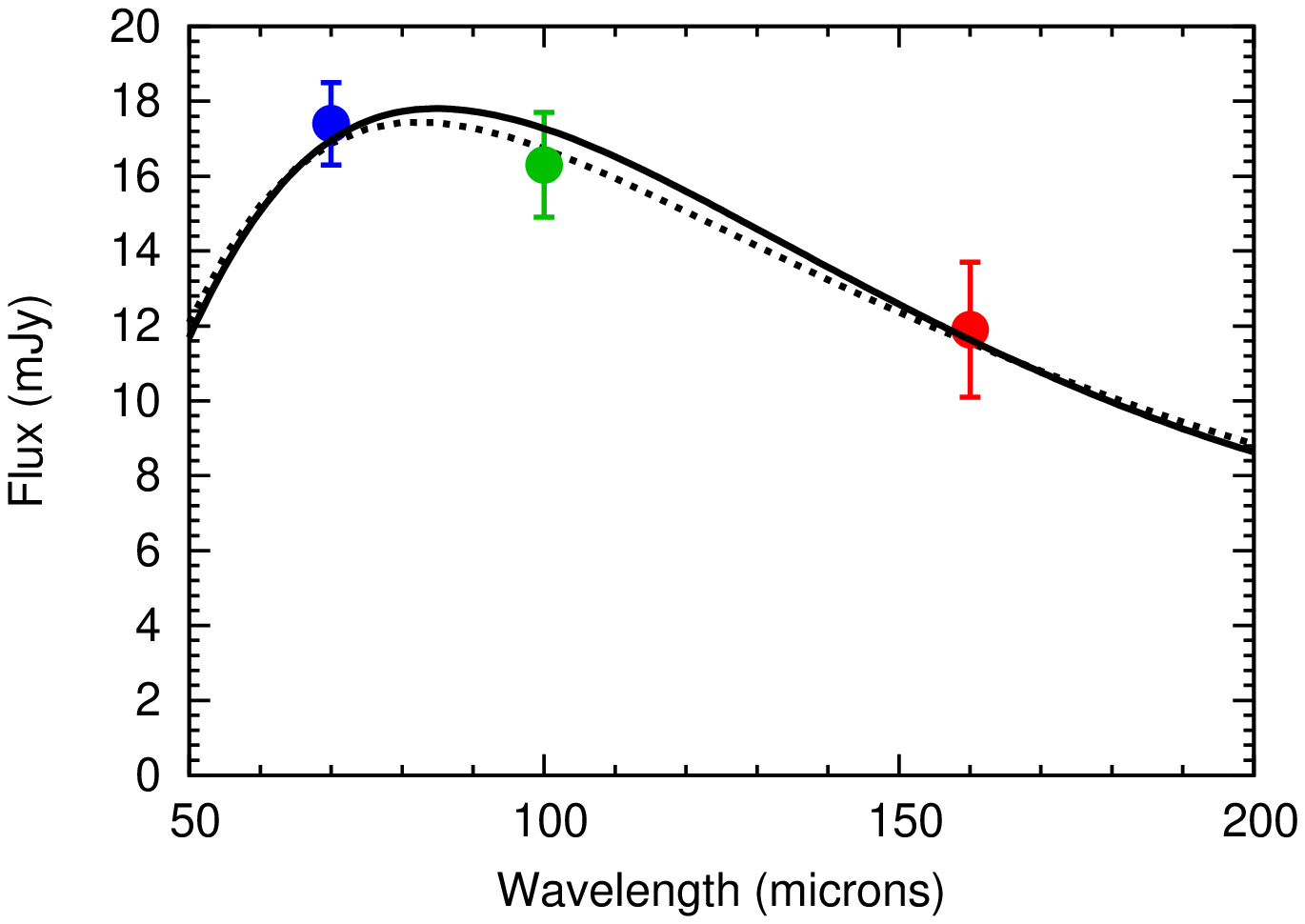}}
\end{center}\vspace*{-4mm}
\caption{Spectral energy distribution of Sedna (left)
and \ek{} (right) in the far-infrared region, based on
Herschel/PACS measurements. Superimposed
are the best-fit TPM (solid lines) and 
STM curves with floating beaming parameter (dashed lines).}
\label{fig:sednaeksed}
\end{figure*}

We found that all three methods yielded the same fluxes \emph{and}
uncertainties for each channel.
The individual analysis of maps for \ek{} also showed consistent
results. Therefore, we accepted the means of all measurements per object
and filter as final fluxes
(see Table~\ref{table:herscheldata}) and used them for thermal modeling. 
We note here that the color
corrections provided in \cite{poglitsch2010} are negligible: it is nearly
or less than $1$\,per cent for \ek{} and for Sedna, it is $6$\,per cent
in the $70\,\mu{\rm m}$ channel and less than $3$\,per cent in the
other longer wavelength channels, so almost less by a magnitude than
the relative photometric uncertainties in all of these cases. 


\section{Thermal properties}
\label{sec:thermal}

The basic physical properties of Sedna and \ek{} were 
estimated by a hybrid standard thermal model 
\citep[STM,][]{lebofsky1986,stansberry2008}
in which the beaming parameter is 
adjustable 
and the asteroid thermophysical model 
\citep[TPM, see][]{lagerros1996,lagerros1997,lagerros1998}.
The absolute magnitudes of the reflected sunlight from these TNOs
are available from the literature
\citep{rabinowitz2007,sheppard2011}. In addition for \ek{}, we 
conservatively increased the formal uncertainty ($0.03$\,mag, based on
MPC data) up to $0.10$\,mag: we took into account the possible omission 
of the phase angle corrections and also added quadratically an 
average TNO lightcurve 
amplitude of $0.088$\,mag \citep[based on][]{duffard2009}. The employed
geometric parameters and absolute magnitudes are summarized in 
Table~\ref{table:auxdata}.

The hybrid STM predicts thermal fluxes from the geometric albedo, 
diameter, and beaming parameter and these fluxes can be computed for 
arbitrary solar and geocentric distances. 
Hybrid STM provides reliable 
estimates only for small phase angles (via a simple form of phase angle 
corrections), although owing to the distances of these objects, this 
estimate is fairly sufficient in our cases. To estimate 
the physical parameters and their respective uncertainties, we used a 
Monte Carlo (MC) approach by varying the fit input around their mean with the 
standard deviation equal to their respective uncertainty. For both targets, 
we used the fixed-$\eta$ approach for the beaming parameter, i.e. 
taking $\eta=1.14\pm0.15$ for each MC step. This mean value of and scatter 
in the beaming factor are taken from \cite{santossanz2012}
and seem to be an acceptable approach for TNOs.
To estimate the phase integral $q$, i.e. the ratio of the Bond to
geometric albedo (i.e. $A=qp_{\rm V}$), we employed an iterative approach. 
First, the phase integral is computed for unity slope parameter 
($G=1$, i.e. $q=0.29+0.68G$), and then refined using eq.~1
of \cite{brucker2009} until convergence. This procedure applied for 
hybrid STM yielded the diameter, geometric albedo, and slope parameter of 
$D=1060\pm100$\,km, $p_{\rm V}=0.290\pm0.061$, and $G=0.42\pm0.04$ for Sedna and
$D=535\pm30$\,km, $p_{\rm V}=0.187\pm0.027$, and $G=0.37\pm0.03$ for \ek{}, 
respectively. 
\refmark{We repeated the similar procedure by allowing the 
beaming parameter $\eta$ to vary. This analysis yielded $\eta=0.95\pm0.43$ for
Sedna, with the corresponding diameter and albedo of 
$D=990\pm95$\,km and $p_{\rm V}=0.336\pm0.072$. For \ek{}, the best-fit 
value of the beaming parameter is somewhat smaller, $0.70\pm0.31$, while the
diameter and albedo values are 
$D=450\pm35$\,km and $p_{\rm V}=0.261\pm0.047$. }

\refmark{In the case of Sedna, we note that
the linear phase coefficient $\beta=0.151\pm0.033$ \citep{rabinowitz2007}
would imply a phase integral of $q=0.89^{+0.55}_{-0.29}$,
assuming the same phase behavior over the whole phase angle range.
Although the phase curve is known for very small domains
\citep[$\alpha\lesssim0.6^\circ$, see also fig.~2 of][]{rabinowitz2007},
this value broadly agrees, within a nearly 1-$\sigma$ difference
from the phase integral of $q=0.59\pm0.03$
as implied by the radiometric albedo and the Brucker formula.}

The results of the TPM estimates were the following. 
For Sedna, we used the rotation period of
$\approx10.27$\,hours \citep{gaudi2005} and assumed an equator-on rotation
and the most favorable solution for the thermal inertia was found 
to be $0.2\,{\rm J}\,{\rm m}^{-2}\,{\rm K}^{-1}\,{\rm s}^{-1/2}$, 
which corresponds to the diameter of $D=995 \pm 80\,{\rm km}$ and 
the geometric albedo of $p_{\rm V}=0.32 \pm 0.06$. 
For \ek{}, we assumed a period of 12\,hours and equator-on rotation
and found that this object also requires a very low thermal 
inertia, $0.1\,{\rm J}\,{\rm m}^{-2}\,{\rm K}^{-1}\,{\rm s}^{-1/2}$.
This may change slightly if the rotation period and the spin vector 
orientation were very different, although all feasible solutions 
put the thermal inertia 
below $1.0\,{\rm J}\,{\rm m}^{-2}\,{\rm K}^{-1}\,{\rm s}^{-1/2}$.
Our best fit yielded a diameter of $D=470^{+35}_{-10}\,{\rm km}$ and a geometric albedo
of $p_{\rm V}=0.25^{+0.02}_{-0.05}$, which do not differ significantly
from the hybrid STM model results \refmark{in which the
beaming parameter was also varied.
We note that here the uncertainties include both the statistical 
errors and the ambiguities in the rotation parameters
\citep[see also][for more details about this modeling]{mueller2011}.}

In Fig.~\ref{fig:sednaeksed}, we displayed the far-infrared SEDs
for these two objects as it is estimated from the hybrid STM and 
TPM fitting and our best-fit data and the floating $\eta$ values. 
We also note that the rotation period of Sedna found by \cite{gaudi2005} 
corresponds to a peak-to-peak amplitude of $0.02$\,mag. 
The small amplitude does not change the 
reliability of the thermal modeling and the 
corresponding shape effects are not relevant to the size determination.


\section{Discussion}
\label{sec:discussion}

We have estimated the sizes and surface albedos 
for the trans-Neptunian objects Sedna and \ek{} using recent observations
of their thermal emission at $70/100/160\,\mu{\rm m}$ with Herschel/PACS. 
On the basis of earlier Spitzer measurements, Sedna had already only an upper limit
to its size estimate \citep{stansberry2008}. 
Our analysis has shown that for Sedna and \ek{}, the respective geometric 
albedos are $p_{\rm V}=0.32\pm0.06$ and $p_{\rm V}=0.25^{+0.02}_{-0.05}$,
thus both objects have brighter surfaces than the average TNO population
\citep{stansberry2008} or SDOs/detached population 
\citep[table~5 in][]{santossanz2012}.
We note that the albedos of Sedna and \ek{} closely match those of 
detached objects in fig.~4a ($p_{\rm V}$ vs. diameter) in \cite{santossanz2012}.
According to \cite{schaller2007} and \cite{brown2011}, 
even with these newly derived parameters,
Sedna lies in the region in which volatiles are expected to be 
retained in the surface (see Fig.~1 in that paper for an
equivalent temperature of $20\pm2$\,K), hence one can 
also expect a brighter surface \citep[see also][]{barucci2005,emery2007}.
Sedna is currently approaching its perihelion. Thus, if the
brightness of the surface were changing owing to the ongoing 
sublimation of ices, it might be detectable in the variation
in the absolute magnitude on a timescale of decades.
In contrast, \ek{} falls in the region
in which \refmark{volatiles should have been lost} (using $48\pm3$\,K for 
equivalent temperature). However, objects of this size can have such a high
albedo if water ice is present on the surface 
\citep[see e.g.][]{barkume2006,ragozzine2009,dumas2011}.
\refmark{The presence of water could be tested by measuring the intrinsic color 
to see whether it is bluish \citep{brown2008}.}
\refmark{As they lack known satellites, we do not know the
masses and hence the surface gravity and escape velocities of these objects
that could place tighter constraints on the surface properties.}


\begin{acknowledgements}
 We gratefully thank the valuable comments and suggestions of the referee,
Josh Emery.
 The work of A.~P., Cs.~K., and N.~Sz. has been supported by the 
ESA grant PECS~98073. A.~P. and Cs.~K. were also supported by the J\'anos 
Bolyai Research Scholarship of the Hungarian Academy of Sciences. 
 Cs.~K. thanks the support of the OTKA grant K101393.
 Part of this work was supported by the German DLR projects number 50 OR 1108,
50 OR 0903, and 50 OR 0904.
 M.~R. acknowledges support from the German Deutsches Zentrum f\"ur Luft- und
Raumfahrt, DLR project number 50OFO~0903.
 M.~M. acknowledges support through the DFG Special Priority Program 1385.
 P.~S.-S. would like to acknowledge financial support by the
Centre National de la Recherche Scientifique (CNRS).
 J.-L.~O. acknowledges the Spanish grants
AYA2008-06202-C03-01, AYA2011-30106-C02-01 and 2007-FQM2998.
\end{acknowledgements}


{}

\end{document}